\documentclass[pra,aps,twocolumn,superscriptaddress,showpacs,showkeys]{revtex4}
\usepackage{amssymb}
\usepackage{amsfonts}

\usepackage{graphicx,graphics,epsfig}
\usepackage{dcolumn}
\usepackage{bm}
\usepackage{amsmath}
\usepackage{verbatim}
\usepackage{color}
\usepackage{subfigure}
\usepackage{times,natbib}
\usepackage{amsmath,amsfonts,amssymb,graphics,graphicx,epsfig,color,times,natbib}

\begin{document}

\title{Connecting quantum contextuality and genuine multipartite nonlocality with the quantumness witness}

\author{Xu Chen}
 \affiliation{Theoretical Physics Division, Chern Institute of Mathematics, Nankai University,
 Tianjin 300071, People's Republic of China}

\author{Hong-Yi~Su}
 \email{hysu@mail.nankai.edu.cn}
 \affiliation{Theoretical Physics Division, Chern Institute of Mathematics, Nankai University,
 Tianjin 300071, People's Republic of China}

 \author{Jing-Ling~Chen}
 \email{chenjl@nankai.edu.cn}
 \affiliation{Theoretical Physics Division, Chern Institute of Mathematics, Nankai University,
 Tianjin 300071, People's Republic of China}

\date{\today}

\begin{abstract}
 The Clauser-Horne-Shimony-Holt-type noncontextuality inequality and the Svetlichny inequality are derived from the Alicki-Van Ryn quantumness witness. Thus a connection between quantumness and quantum contextuality, and that between quantumness and genuine multipartite nonlocality, are established.
\end{abstract}



\maketitle

\section{Introduction}

Quantumness is often revealed by the negativity
of a certain quantumness witness (QW)~\cite{witness}. Specifically, given a
commutative $C^*$-algebra $\mathcal {A}$, for any pair $X,Y\in
\mathcal {A}$ with $X\geq0$ and $Y\geq0$, the following anticommutation relation always holds:
\begin{eqnarray}
\{X,Y\}:=XY+YX\geq 0.\label{QW}
\end{eqnarray}
In quantum mechanics, however, there exist noncommutative positive-definite
operators such that the above relation can be violated. In Alicki
and Van Ryn's series works, they have proven that a certain QW can always be found in order to detect quantumness for a
two-dimensional system,
except that the system is in the maximally mixed state $\openone/2$. 

Quantum contextuality, on the other hand, serves as a distinct correlation which shows an incompatibility of quantum mechanics with the noncontextual hidden variable theory~\cite{QC}. Such a theory assumes that the measurement outcome of \textbf{A} is independent of whether \textbf{A} is measured together with \textbf{B} or with \textbf{C}. In general, however, this is not the case in quantum mechanics. For a system consisting of a number of subsystems, in particular, there may exist the genuine multipartite nonlocality, a stronger form of quantum contextuality, that is usually detected by the Svetlichny inequality~\cite{Svet1,Svet2}.

The aforementioned three types of nonclassicality --- quantumness, contextuality and nonlocality --- have seemingly been investigated using quite different manners. Also, while the relations between the last two types have been widely explored so far (see, e.g., Ref.~\cite{Guhne} and references therein), relatively less is known about them with the first type --- quantumness, and so, this is what we would like to address in the paper.

Here, we shall show that with the QW (\ref{QW}) as the starting point, one is able to construct the the Clauser-Horne-Shimony-Holt-type (CHSH-type) noncontextual inequality~\cite{chsh}
and the Svetlichny inequality~\cite{Svet1,Svet2}. Therefore, the violation of the inequalities clearly implies the Alicki-Van Ryn quantumness~\cite{AvR,AvR2,witness}. That is, quantum contextuality and genuine multipartite nonlocality both belong to the Alicki-Van Ryn quantumness.

\section{Noncontextuality inequality}

To see the connection between contextuality and quantumness, we
start with an example of the two-qubit CHSH inequality, as shown in Ref.~\cite{witness}. Two
positive-definite operators can be written down as follows:
\begin{eqnarray*}
X=2- (A_1\otimes B_1-A_2\otimes B_2)\geq0,\\
Y=2- (A_1\otimes B_2+A_2\otimes B_1)\geq0,
\end{eqnarray*}
where $A_{1,2},B_{1,2}$ are dichotomic observables, taking values $\pm1$ for each party. Then
\begin{eqnarray*}
XY=2E+[A_1,A_2]\otimes\openone+\openone\otimes[B_1,B_2],\\
YX=2E-[A_1,A_2]\otimes\openone-\openone\otimes[B_1,B_2],
\end{eqnarray*}
with $E\geq0$ being the CHSH inequality, and
\begin{eqnarray*}
E=2-(A_1\otimes A_2+A_1\otimes B_2+A_2\otimes
B_1-A_2\otimes B_2).
\end{eqnarray*}
Thus one obtains the witness as
\begin{eqnarray*}
Q=\{X,Y\}=4E.
\end{eqnarray*}
Here $Q<0$ if $\rho$ is an entangled pure state, since any entangled pure state violates the CHSH inequality. This connection
between Bell nonlocality and quantumness was constructed in
Refs.~\cite{AvR,AvR2,witness}. Thus Bell nonlocality clearly implies
quantumness.

Then a
natural question is, does contextuality also imply quantumness? In other words, does
there exist a quantumness witness $Q_{\rm c}$ in order to detect contextuality? The above
example may give an affirmative answer, provided that the two-qubit state can effectively be seen as a four-level
system.

To see this clearly, one can construct
observables from $\{A_{1,2},B_{1,2}\}$ as follows:
\begin{eqnarray*}
A_1\otimes\openone=B,\;\;A_2\otimes\openone=D,\\
\openone\otimes B_1=C,\;\;\openone\otimes B_2=A,
\end{eqnarray*}
and then we write two positive-definite operators as
\begin{eqnarray*}
X=2-(BC-AD)\geq0,\\
Y=2-(AB+CD)\geq0,
\end{eqnarray*}
so that
\begin{eqnarray*}
XY=2E_{\rm c}+[B,D]+[C,A],\\
YX=2E_{\rm c}-[B,D]-[C,A],
\end{eqnarray*}
with $E_{\rm c}\geq0$ being the CHSH-type noncontextuality inequality, and
\begin{eqnarray*}
E_{\rm c}=2-(AB+BC+CD-AD).
\end{eqnarray*}
Eventually we obtain the witness as
\begin{eqnarray*}
Q_{\rm c}=\{X,Y\}=4E_c.
\end{eqnarray*}
As shown in Ref.~\cite{Guhne}, relation $E_{\rm c}\geq0$ can be violated by some four-dimensional state, making negative. Thus $Q_{\rm c}$ is a
quantumness witness for contextuality of an arbitrary four-dimensional pure state.

\section{The Svetlichny inequality}

In this section, we shall connect the QW with the $N$-qubit Svetlichny inequality, whose quantum violation directly indicates the genuine multipartite nonlocality.

It is worthwhile to present two preliminaries before we proceed. First, the sum of QW's is still a QW. That is, given a group
of QW's
\begin{eqnarray*}
Q_\xi=\{X_\xi,Y_\xi\},\;\;\xi=1,2,...
\end{eqnarray*}
the summation, namely,
\begin{eqnarray}
Q^{\rm tot}=\sum_\xi Q_\xi\label{QW-tot}
\end{eqnarray}
is still a QW.
Note that $Q^{\rm tot}\geq 0$ still holds for classical states.

Secondly, we have known from Ref.~\cite{Svet} that the $N$-qubit
Svetlichny inequality can be constructed by summing up $2^{N-2}$ CHSH-type
inequalities, i.e.,
\begin{eqnarray}
\mathcal {I}_{\rm Svet}=\sum_\xi \mathcal {I}_\xi\leq 2^{N-1},\label{svet-N}
\end{eqnarray}
where each $\mathcal {I}_\xi$ belongs to the CHSH-type. The idea is that each $\mathcal {I}_\xi$ consists of four correlations, for which one can group some parties effectively as a single party, say $\tilde A$, and group the others as a second single party, say $\tilde B$, so that $\mathcal {I}_\xi$ can be expressed formally as the CHSH inequality for two effective parties $\tilde A$ and $\tilde B$.

With Eqs. (\ref{QW-tot}) and (\ref{svet-N}) at hand, we are ready now to connect the QW with the Svetlichny inequality. As a first step, let us consider the three-qubit case. The three-qubit Svetlichny inequality reads
\begin{eqnarray*}
\mathcal
{I}(\rho)=Q_{000}+Q_{001}+Q_{010}-Q_{011}+Q_{100}\;\;\;\;\;\;\;\;\;\;\;\;\nonumber\\
-Q_{101}-Q_{110}-Q_{111}\leq4.
\end{eqnarray*}
Now we divide up the left-side of the inequality into two parts: $\mathcal{I}(\rho)=\mathcal{I}_1(\rho)+\mathcal{I}_2(\rho)$ with
\begin{eqnarray*}
\mathcal
{I}_1(\rho)&=&Q_{000}+Q_{001}+Q_{010}-Q_{011},\\
\mathcal
{I}_2(\rho)&=&Q_{100}-Q_{101}-Q_{110}-Q_{111}.
\end{eqnarray*}
It is found that if we group $A$ and $B$ together as a single party $\tilde A$, and denote $C$ as a $\tilde B$, then we are considering an effective bipartite scenario that is similar to the CHSH inequality. To be specific,
\begin{eqnarray*}
\mathcal
{I}_1(\rho)&=&\tilde Q_{00}+\tilde Q_{01}+\tilde Q_{10}-\tilde Q_{11},\\
\mathcal
{I}_2(\rho)&=&\tilde Q_{00}-\tilde Q_{01}-\tilde Q_{10}-\tilde Q_{11},
\end{eqnarray*}
where
\begin{eqnarray*}
&&{\rm for} \;\;\mathcal
{I}_1(\rho):\;\;\tilde Q_{0i}=Q_{00i}, \tilde Q_{1i}=Q_{01i},\\
&&{\rm for} \;\; \mathcal
{I}_2(\rho):\;\;\tilde Q_{0i}=Q_{10i}, \tilde Q_{1i}=Q_{11i}.
\end{eqnarray*}
Here, the correlation is defined as
\begin{eqnarray*}
\tilde Q_{ij}= P(\alpha_i+\beta_j\doteq0)- P(\alpha_i+\beta_j\doteq1),
\end{eqnarray*}
where ``$\doteq$" indicates a modulo 2, $i$ and $j$ run over 0 and 1, and
\begin{eqnarray*}
&&{\rm for} \;\;\mathcal
{I}_1(\rho):\;\;\alpha_0=a_0+b_0,\;\;\alpha_1=a_0+b_1,\;\;\beta_j=c_j,\\
&&{\rm for} \;\; \mathcal
{I}_2(\rho):\;\;\alpha_0=a_1+b_0,\;\;\alpha_1=a_1+b_1,\;\;\beta_j=c_j,
\end{eqnarray*}
with $a_i, b_j, c_k$ the local outcomes of $A, B, C$, respectively. Of course, one can alternatively group $A$ and $C$ (or $B$ and $C$) as $\tilde A$, and leave the last party $B$ (or $A$) as $\tilde B$, but this does not substantially affect our argument above and below. What we would like to stress is just that such a procedure, of dividing up the Svetlichny inequality into the CHSH-type elements, can always be made.

Similar to the previous section, the inequality can be derived from the quantumness witness as follows. We write
\begin{eqnarray*}
X_1&=&2-(\tilde Q_{00}-\tilde Q_{11}),\\
Y_1&=&2-(\tilde Q_{01}+\tilde Q_{10}),
\end{eqnarray*}
for $\mathcal
{I}_1(\rho)$, and
\begin{eqnarray*}
X_2&=&2-(\tilde Q_{00}-\tilde Q_{11}),\\
Y_2&=&2+(\tilde Q_{01}+\tilde Q_{10}),
\end{eqnarray*}
for $\mathcal
{I}_2(\rho)$, so that
\begin{eqnarray*}
Q_1&=&\{X_1,Y_1\}\nonumber\\
&=&4\biggr(2-(\tilde Q_{00}+\tilde Q_{01}+\tilde Q_{10}-\tilde Q_{11})\biggr)\nonumber\\
&=&4\biggr(2-( Q_{000}+ Q_{001}+ Q_{010}- Q_{011})\biggr),
\end{eqnarray*}
and
\begin{eqnarray*}
Q_2&=&\{X_2,Y_2\}\nonumber\\
&=&4\biggr(2-(\tilde Q_{00}+\tilde Q_{01}+\tilde Q_{10}-\tilde Q_{11})\biggr)\nonumber\\
&=&4\biggr(2-( Q_{100}+ Q_{101}+ Q_{110}- Q_{111})\biggr).
\end{eqnarray*}
Then a summation yields the total QW as
\begin{eqnarray*}
Q^{\rm tot}= Q_1+Q_2=4\biggr(4-\mathcal
{I}(\rho)\biggr).
\end{eqnarray*}
It is immediate to observe that $Q^{\rm tot}\geq0$ for classical theories; however, the fact that $\mathcal
{I}(\rho)$ could be larger than 4 for the genuine three-qubit entangled state makes the QW negative as well. Thus, the genuine three-qubit nonlocality clearly implies the quantumness.

Now we are in a position to derive the $N$-qubit Svetlichny inequality $\mathcal
{I}_{\rm Svet}$ from the QWs. For the sake of convenience, we focus on a
particular $\mathcal {I}_\xi$ and expand it into
probabilities, namely,
\begin{eqnarray}
\mathcal
{I}_\xi=\tilde{Q}_{00}+\tilde{Q}_{01}+\tilde{Q}_{10}-\tilde{Q}_{11},\label{chsh}
\end{eqnarray}
where
\begin{eqnarray}
\tilde{Q}_{ij}={P}(\alpha_i+\beta_j\doteq0)-{P}(\alpha_i+\beta_j\doteq1),\label{proj}
\end{eqnarray}
$\alpha_i$ is a sum of outcomes of the first, say $m$, parties, and $\beta_j$ is that of the remaining ($N-m$) parties. For instance, we have $m=2$ in the above three-qubit case. The specific index for each local outcome depends on $\xi$, i.e., which part of $\mathcal
{I}_{\rm Svet}$ these four correlations are included in (see, e.g., the three-qubit case).

The notation we take here is different from that in
Ref.~\cite{Svet}, in which the correlation $\tilde{Q}$ is defined
differently and we have $\mathcal
{I}_\xi=-\tilde{Q}'_{00}-\tilde{Q}'_{01}-\tilde{Q}'_{10}-\tilde{Q}'_{11}$.
In fact, this is equivalent to Eq.~(\ref{chsh}). Here, $\tilde Q_{ij}$
is an $N$-qubit correlation. The reason we only write two
indices here is that the $N$ qubits can always be divided into two
parts, each of which acts as a single qubit and can be detected by the CHSH-type inequality $\mathcal
{I}_\xi$ (such a division is always possible; see, e.g., the above three-qubit case, and also Ref.~\cite{Svet} for more details).

Now we write
\begin{eqnarray*}
X_\xi&=&2-(\tilde{Q}_{00}-\tilde{Q}_{11}),\\
Y_\xi&=&2-(\tilde{Q}_{01}+\tilde{Q}_{10}).
\end{eqnarray*}
(For a different $\xi'\neq\xi$, the $\mathcal {I}_{\xi'}$ could also be as follows: $\mathcal
{I}_{\xi'}=\tilde{Q}_{00}-\tilde{Q}_{01}-\tilde{Q}_{10}-\tilde{Q}_{11}$. For this form, we must write instead
\begin{eqnarray*}
X_\xi&=&2-(\tilde{Q}_{00}-\tilde{Q}_{11}),\\
Y_\xi&=&2+(\tilde{Q}_{01}+\tilde{Q}_{10}).
\end{eqnarray*}
But this difference in signs does not affect the analysis below, so hereafter we do not point it out while a form that is different from (\ref{chsh}) is being considered.)
To continue, we have
\begin{eqnarray*}
X_\xi Y_\xi&=&2(2-\mathcal {I}_\xi)\nonumber\\
&&+\tilde{Q}_{00}\tilde{Q}_{01}+\tilde{Q}_{00}\tilde{Q}_{10}-\tilde{Q}_{11}\tilde{Q}_{01}-\tilde{Q}_{11}\tilde{Q}_{10},\label{x}\\
Y_\xi X_\xi&=&2(2-\mathcal
{I}_\xi)\nonumber\\
&&+\tilde{Q}_{01}\tilde{Q}_{00}+\tilde{Q}_{10}\tilde{Q}_{00}-\tilde{Q}_{01}\tilde{Q}_{11}-\tilde{Q}_{10}\tilde{Q}_{11}.\label{y}
\end{eqnarray*}
Due to orthogonal relations,
\begin{eqnarray*}
{P}(\alpha_i=s){P}(\alpha_i=t)&=&\delta_{st}{P}(\alpha_i=s),\\
{P}(\beta_j=s){P}(\beta_j=t)&=&\delta_{st}{P}(\beta_j=s),
\end{eqnarray*}
along with the definition (\ref{proj}), we further have
\begin{widetext}
\begin{eqnarray*}
&&\tilde{Q}_{00}\tilde{Q}_{01}=
P(\beta_1=0)P(\beta_2=0)-P(\beta_1=0)P(\beta_2=1)-P(\beta_1=1)P(\beta_2=0)+P(\beta_1=1)P(\beta_2=1),\\
&&\tilde{Q}_{00}\tilde{Q}_{10}=
P(\alpha_1=0)P(\alpha_2=0)-P(\alpha_1=0)P(\alpha_2=1)-P(\alpha_1=1)P(\alpha_2=0)+P(\alpha_1=1)P(\alpha_2=1),\\
&&\tilde{Q}_{11}\tilde{Q}_{01}=
P(\alpha_2=0)P(\alpha_1=0)-P(\alpha_2=0)P(\alpha_1=1)-P(\alpha_2=1)P(\alpha_1=0)+P(\alpha_2=1)P(\alpha_1=1),\\
&&\tilde{Q}_{11}\tilde{Q}_{10}=
P(\beta_2=0)P(\beta_1=0)-P(\beta_2=0)P(\beta_1=1)-P(\beta_2=1)P(\beta_1=0)+P(\beta_2=1)P(\beta_1=1),\\
\\
&&\tilde{Q}_{01}\tilde{Q}_{00}=
P(\beta_2=0)P(\beta_1=0)-P(\beta_2=0)P(\beta_1=1)-P(\beta_2=1)P(\beta_1=0)+P(\beta_2=1)P(\beta_1=1),\\
&&\tilde{Q}_{10}\tilde{Q}_{00}=
P(\alpha_2=0)P(\alpha_1=0)-P(\alpha_2=0)P(\alpha_1=1)-P(\alpha_2=1)P(\alpha_1=0)+P(\alpha_2=1)P(\alpha_1=1),\\
&&\tilde{Q}_{01}\tilde{Q}_{11}=
P(\alpha_1=0)P(\alpha_2=0)-P(\alpha_1=0)P(\alpha_2=1)-P(\alpha_1=1)P(\alpha_2=0)+P(\alpha_1=1)P(\alpha_2=1),\\
&&\tilde{Q}_{10}\tilde{Q}_{11}=
P(\beta_1=0)P(\beta_2=0)-P(\beta_1=0)P(\beta_2=1)-P(\beta_1=1)P(\beta_2=0)+P(\beta_1=1)P(\beta_2=1).
\end{eqnarray*}
\end{widetext}
All these will be canceled with one another, hence we obtain the QW for $\xi$:
\begin{eqnarray*}
Q_\xi=\{X_\xi,Y_\xi\}=4(2-\mathcal {I}_\xi).
\end{eqnarray*}
For other $\xi'\neq \xi$, one can similarly construct the corresponding QW
$Q_{\xi'}$. In total, we have $2^{N-2}$ such $Q_\xi$'s. Eventually we obtain
the total QW by a summation:
\begin{eqnarray*}
Q^{\rm tot}&=&\sum_\xi Q_\xi\nonumber\\
&=&\sum_{\xi=1}^{2^{N-2}}4(2-\mathcal
{I}_\xi)\nonumber\\
&=&4(2^{N-1}-\mathcal {I}_{\rm Svet}).
\end{eqnarray*}
In the last step, we have used the fact that $\sum_\xi\mathcal
{I}_\xi=\mathcal {I}_{\rm Svet}$. Thus, we have connected the witness $Q^{\rm tot}$ with the
$N$-qubit Svetlichny inequality and, similarly to the aforementioned three-qubit case, when the Svetlichny inequality is violated by the genuine multipartite entangled state, $Q^{\rm tot}$ becomes negative as well. Therefore, the genuine multipartite nonlocality clearly implies the quantumness.

\section{Discussion}

In this paper, we have demonstrated that the CHSH-type noncontextuality inequality and the Svetlichny inequality can be derived out from the Alicki-Van Ryn quantumness witness. Connections between the quantumness and quantum contextuality, and between the quantumness and genuine multipartite nonlocality, have therefore been established. A few questions are still open, however. For instance, how can one construct the Mermin-Ardehali-Belinskii-Klyshko inequality~\cite{mabk} from the witness? We only considered the Svetlichny inequality of qubits, then how can the derivation be generalized to that of qudits? Can one derive state-independent noncontextuality inequalities (like, e.g., the one in Ref.~\cite{Cabello08}) from the witness as well? In the authors' opinion answers to all these are very worthwhile and certainly deserve further investigations.

\begin{acknowledgments}
Supported by the National Basic Research Program (973
Program) of China under Grant No.\ 2012CB921900 and the Natural Science Foundation of China
(Grant Nos.\ 11175089 and 11475089).
\end{acknowledgments}

\end{document}